\documentclass[preprint,12pt]{elsarticle}

\usepackage{amssymb}

\journal{Physica A}

\newcommand{\DF}[2]{{\displaystyle\frac{#1}{#2}}}

\begin{document}

\begin{frontmatter}



\title{Integrating neighborhoods in the evaluation of fitness promotes cooperation in the spatial prisoner's dilemma game}


\author{Zhen Wang$^{a}$, Wen-Bo Du$^{b,c,\ast}$, Xian-Bin Cao$^{b,\ast}$, Lian-Zhong Zhang$^{a,\ast}$}

\address{$^a$ School of Physics, Nankai University, Tianjin, 300071 P.R. China\\
$^b$ School of Electronic and Information Engineering, Beihang
University, Beijing, 100083, P.R.China\\
$^c$ School of Computer Science and Technology, University of Science and Technology of China, Hefei, 230026, P. R. China\\
{E-mail:wenbodu@mail.ustc.edu.cn,xbcao@buaa.edu.cn,zhanglz@nankai.edu.cn}}

\begin{abstract}
A fundamental question of human society is the evolution of
cooperation. Many previous studies explored this question
 via setting spatial background, where players obtain their
 payoffs by playing game with their nearest neighbors. Another
undoubted fact is that environment plays an important role in the
individual development. Inspired by these phenomena, we reconsider
the definition of individual fitness which integrates the
environment, denoted by the average payoff of all individual
neighbors, with the traditional individual payoffs by introducing a
selection parameter $u$. Tuning $u$ equal to zero returns the
traditional version, while increasing $u$ bears the influence of
environment. We find that considering the environment, {\it i.e.},
integrating neighborhoods in the evaluation of fitness, promotes
cooperation. If we enhance the value of $u$, the invasion of
defection could be resisted better. We also provide quantitative
explanations and complete phase diagrams presenting the influence of
environment on the evolution of cooperation. Finally, the
universality of this mechanism is testified for different
neighborhood sizes, different topology structures and different game
models. Our work may shed a light on the emergence and persistence
of cooperation in our life.\\
\end{abstract}


\end{frontmatter}



\section{Introduction}

The evolution of cooperation among unrelated individuals has become
a major challenge in the biology and evolution research since the
altruism and unselfish seems incompatible with Darwinian selection.
Nevertheless, cooperation can be abundantly found in animal and
human societies. In order to interpret these universal cooperative
phenomena, evolutionary game theory has become a useful tool to
study this puzzling dilemma \cite{1,2,3}. Of particular renown is
the prisoner's dilemma game, which as a paradigm illustrates the
social conflict between altruistic and selfish behaviors and has
attracted much attention in both theoretical and experimental
studies \cite{4}. In the original form of the game, there exist two
players, and they have to simultaneously decide whether to cooperate
(C) or to defect (D). Naturally, their payoffs depend on their
decisions: they can get $R$ ($P$) while mutual cooperation
(defection); if a defector meets a cooperator, the former can obtain
a maximum individual payoff $T$ and the latter can only get a
minimal payoff $S$. The ranking of the four payoffs must satisfy $T>
R > P > S$ and $2R > T + S$. Since the defector always outperforms
the cooperator, irrespective of the opponent's choice, two
players will inevitably fall into the mutual defection state. Over
the past decades, many mechanisms have been proposed to overcome
this dilemma, such as kin selection \cite{5}, direct and indirect
reciprocity \cite{6,7}, voluntary participation \cite{8} and spatial
structure \cite{9,10}. Among these achievements, the most prominent
success is obtained by the spatial extension.

In the pioneer work by Nowak and May \cite{10}, players were
arranged on the vertex of a square lattice, and their payoffs were
gathered from playing the game with their nearest neighbors. Then players
were allowed to adopt the strategy of their neighbors, provided
their payoff was higher. It was shown that the emergence and
sustainment of cooperation could be greatly improved. Along this way,
much effort has been given to discover new mechanisms that can
sustain stable cooperation. They include the interplay between
evolutionary games and network structure (regular networks
\cite{11,12}, small world networks \cite{13,14,15} and scale-free
networks \cite{16,17,18}), environmental noise \cite{19,20,21},
additional noise introducing into payoffs \cite{22,23}, teaching
activity \cite{24,25,26}, the mobility of players \cite{27,28},
asymmetric payoff \cite{29,30}, differences in evolutionary time
scales \cite{31}, aspiring to the fittest \cite{32,33}. For recent
surveys of this field, one can refer to two extensive reviews
\cite{34,35}.

In the present work, we mainly focus on the research of individual
fitness during the process of evolution. For the majority of previous
researches, a player's fitness in the spatial structure
equals the current accumulated payoff collected from its direct
neighbors. Indeed, the meaning of fitness is often related to the
reproduction ability \cite{36,37} and some works exhibiting the
relationships between payoff and fitness are also presented. For
example, Ohtsuki {\it et.al.} defined fitness as the accumulated
payoff with a background noise \cite{38}; Szolnoki {\it et.al.}
featured fitness as mixture of accumulated payoff and normalized
payoff \cite{39}; Jia {\it et.al.} drew fitness as the accumulated
payoff with a fluctuation coefficient \cite{40}; and other
researchers described fitness as combination of current payoff and
past payoffs \cite{41,42,43}. Inspired by their instructive
suggestions, we propose a new method to define individual fitness where
neighborhoods are considered.

As is known, environment plays an indispensable role in the
real life and it can usually affect individual development. For
example, people in the elite teams are more likely to learn
technical ability from their excellent teammates to  enhance
competitiveness of themselves, animals in the strong groups are more
easily to prey or resist the predators. Therefore, it is natural
to take environment presented by the average payoff of all individual neighbors into account when talking about individual
fitness. Here we redefine the individual fitness via integrating its current accumulated payoff with average payoff of all individual neighbors, and find that the evolution of cooperation can be remarkably promoted.

The paper is organized as follows. In the next section, we will first describe
evolutionary prisoner's dilemma game and the adjusted
definition of fitness. Subsequently, the main simulation results
and discussions will be presented in Section 3. And the conclusion is summarized
in the last section.

\section{The Model}

\begin{figure}[htbp]
\begin{center}
\scalebox{0.55}[0.55]{\includegraphics{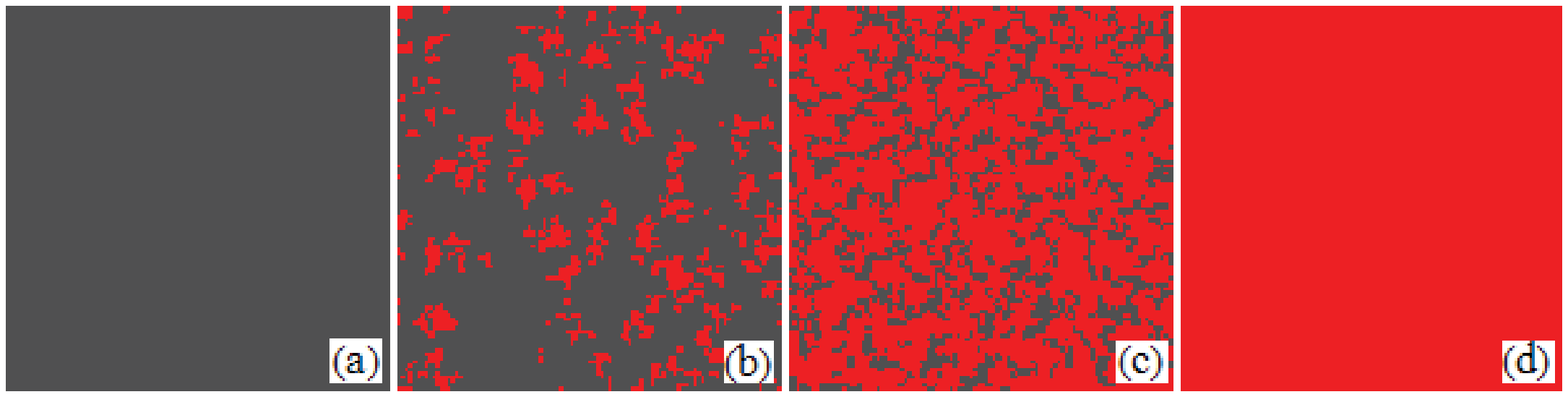}}
\caption{(\textit{color online}) Characteristic snapshots of
cooperators (red) and defectors (gray) for different values of the
 parameter $u$. From (a) to (d) $u = 0$, $0.2$, $0.4$ and
$0.6$, respectively. Depicted results were obtained for $b=1.20$ and
$K=0.1$ on a $100 \times 100$ square lattice.}\end{center}
\end{figure}

For simplicity, but without loss of generality, we consider an evolutionary
prisoner's dilemma game with the rescaled payoff matrix: $T=b$, $R=1$ and $P=S=0$,
where the parameter $b$ ($1< b < 2$)
represents the temptation to choose defection. Initially, cooperators and
defectors are randomly distributed on a square $L \times L$ square lattice with
equal probability. At each time step, player $i$ firstly acquires its payoff $P_{i}$
by playing game with its direct neighbors. Next, the payoffs $P_{j}$ of all the
neighbors of player $i$ can be obtained by means of the same way as player $i$, and
the environment could be characterized by the average value of all neighbors' payoffs $P_{j}$
, that is,

\begin{equation}
\label{abcc} \overline{P}=\frac{\sum_{j=1}^{k_i} {P_j}}{{k_i}} ,
\end{equation}

where the sum runs over all the neighbors of player $i$, and $k_i$ denotes the
neighbor number of player $i$. Then we evaluate the
fitness of player $i$ according to the following expression:

\begin{equation}
\label{abcc} f_{i}=(1-u)\times{P_i}+ u\times{\overline{P}},
\end{equation}

where $0 \leq  u \leq 1$ is a tunable parameter. Obviously, when $u=0$, the fitness mathematically
equals the accumulated payoff $P_i$, which does not consider the influence
of environment and goes back to the traditional version \cite{19,24,32}. Interestingly,
as $u$ increases, the environment will
hold larger proportion. After each step of the game, players
synchronously update their strategies. When player $i$ is to update
its strategy, it will randomly select a neighbor $j$ and adopt its strategy
with a probability $W_{(i \rightarrow j)}$ depending on the fitness difference. namely

\begin{equation}
\label{abcc} W_{(i \rightarrow
j)}=\DF{1}{1+exp[{(f_i-f_j)}/{\it K}]},
\end{equation}

where $0 < {\it K} < +\infty$ characterizes the environmental noise,
including irrationality and errors. The effect of noise
${\it K}$  has been well studied in previous literatures\cite{19,32,33,46}.

\section{Simulation Results and Discussion}

\begin{figure}[htbp]
\begin{center}
\scalebox{0.2}[0.2]{\includegraphics{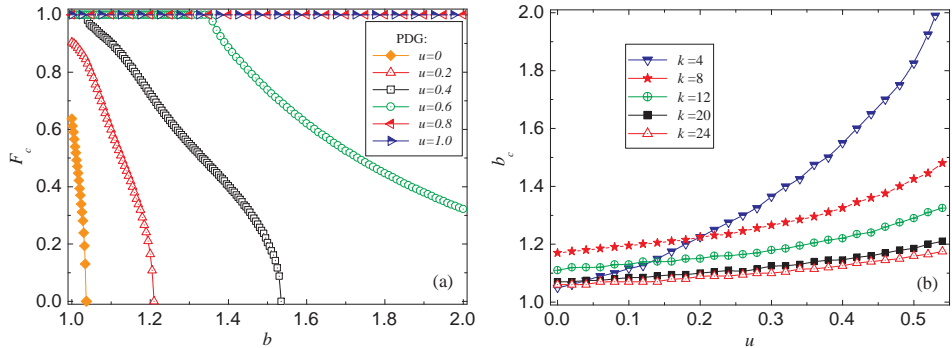}}
\caption{(\textit{color online}) (a):Frequency of cooperators ${\it
F_{\rm c}}$ in dependence on the parameter $b$ for different values
of the  parameter $u$. Note that comparing with the traditional
version ({\it i.e.,} $u=0$), the cooperation can be greatly improved
in our version. (b): Critical threshold values of $b=b_{c}$, marking
extinction of cooperators, in dependence on the parameter $u$ for
different neighborhood sizes. Note that the value of $b_c$ will
monotonously increase with the the increase of $u$, irrespective of
the neighborhood size. Due to the scale of the parameter $b (1 < b <
2)$, $b_{c}>2$ is canceled. Depicted results were obtained for
$K=0.1$.}
\end{center}
\end{figure}

Simulations below were carried out on a $100 \times 100$ square
lattice. The key quantity frequency of cooperators ${\it F_{c}}$ was determined
within $10^4$ full Monte Carlo steps after sufficiently long
transients were discarded. Moreover, each data were averaged over up to 20 independent runs
for each set of parameter values in order to assure suitable
accuracy.

We start by visually inspecting the spatial patterns formed by
cooperators and defectors for different values of the adjusting
parameter $u$. Figure 1 shows the results obtained by $b=1.20$ and
${\it K}=0.1$. For $u=0$ (Fig.1(a)), the model returns
the traditional model and the system falls into the pure-defector state
when $b$ is not very large; for $\mu=0.2$ (Fig.1(b)), a
small fraction of cooperators can sustain by forming the small yet compact
clusters to resist the invasion of defectors; for $u=0.4$ (Fig.1(c)),
cooperators outperform defectors and they can form
strong and large clusters to survive and prevail; for $u=0.6$ (Fig.1(d)),
the system will be entirely dominated by cooperation strategy. These results reveal
that considering the environment, {\it i.e.}, integrating neighborhoods in the evaluation of fitness,
can greatly enhance the evolution of cooperation.

\begin{figure}[htbp]
\begin{center}
\scalebox{0.5}[0.5]{\includegraphics{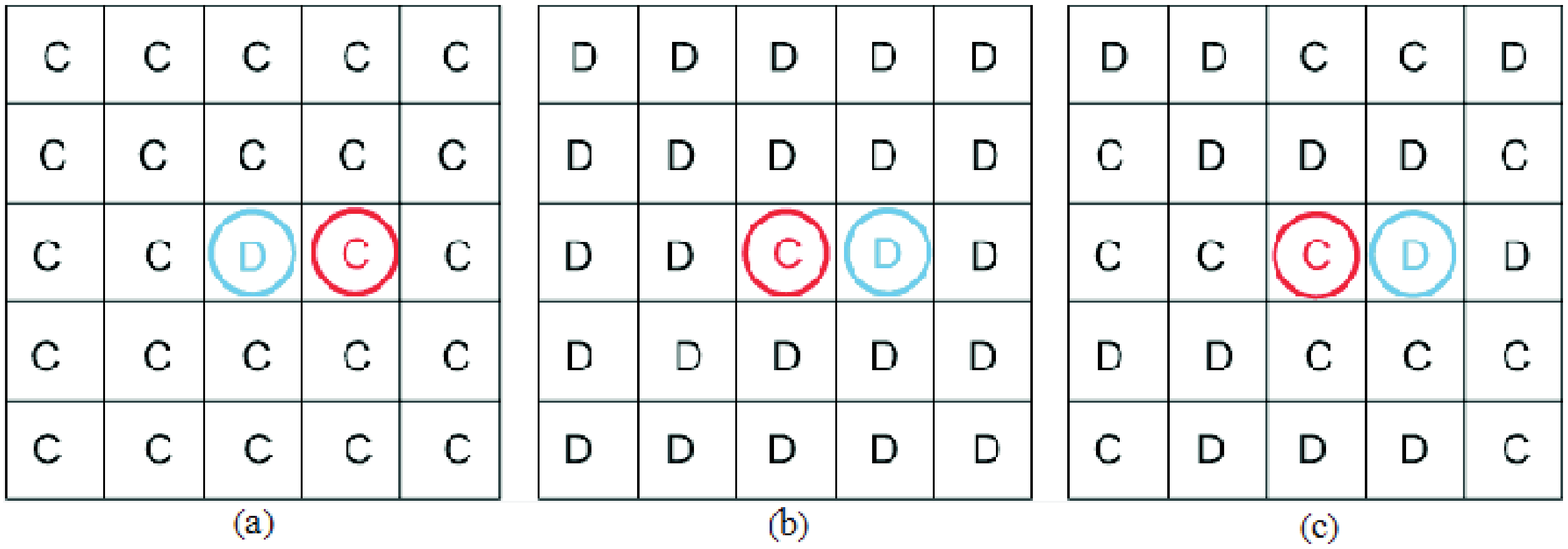}}
\caption{(\textit{color online}) Three typical sub-patterns in the
spatial game: (a) a defector surrounded by cooperators; (b) a
cooperator surrounded by defectors; (c) cooperators and defectors
are evenly mixed.}\label{1}
\end{center}
\end{figure}

To quantify the effect of $u$ more precisely, we examine the
relationship between the frequency of cooperators ${\it F_c}$ and the temptation to
defect $b$ for different values of $u$. As is shown in Fig.2(a),
cooperation monotonously decreases with the increase
of $b$ no matter under what value of $u$. Interestingly, it is evident that $u$
plays a crucial role in the evolution of cooperation: the larger the value of $u$,
the higher the cooperation level. In particular, when $u$ is considerably large
($u \geq 0.8$), the defectors still go extinct even if defectors possess
great advantage over cooperators ($b=2.0$), which indicates that the
effect of $u$ ({\it i.e.}, integrating neighborhoods in the evaluation of fitness)
on facilitating cooperation is very conspicuous. Moreover, it is also worth noting
how the critical value $b_c$, marking the extinction of
cooperators, varies in dependence on the parameter $u$ for different neighborhood sizes.
Fig.2(b) demonstrates clearly the existence of an optimal neighborhood size when the value
of $u$ is small, that is, cooperation can be better promoted in an intermediate neighborhood
size, which is in agreement with the result of traditional version \cite{44}. Whereas,
as the value of $u$ increasing this trend will change, and small neighborhood size will be
more beneficial for the evolution of cooperation. More importantly, we can observe that
although the survival space for cooperators is becoming smaller and smaller, which could be
predicted from the mean-field approximation \cite{34,45}, the
threshold value of cooperators $b_c$ will monotonously enhance with the increasing of $u$,
irrespective of the neighborhood sizes. This suggests that such a facilitation effect on
cooperation is robust to different neighborhood sizes.

\begin{figure}[htbp]
\begin{center}
\scalebox{0.25}[0.25]{\includegraphics{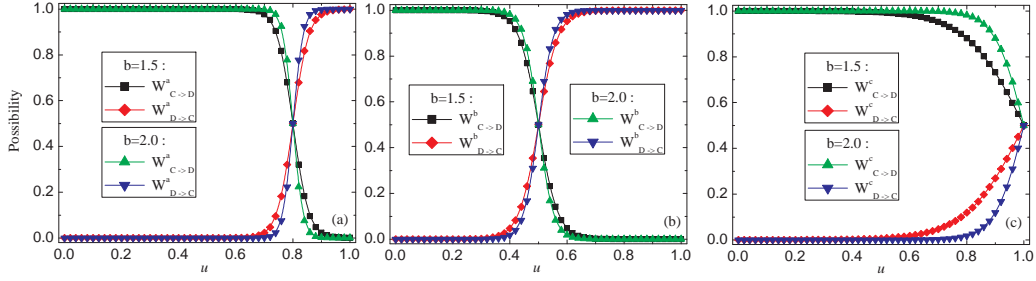}} \caption{(\textit{color online})
The transition possibility between the circled cooperator
and defector in Fig.3. (a) corresponds to the pattern of Fig.3(a);
(b) corresponds to the pattern of Fig.3(b); (c) corresponds to the pattern of Fig.3(c).
Depicted results were obtained for $K=0.1$.
}
\end{center}
\end{figure}

Next we will analyze the underlying mechanism for this kind of
promotion effect. Figure 3 shows three typical sub-patterns on
square lattices, pattern (a): a defector is surrounded by cooperators; pattern (b):  a cooperator
encircled by defectors; pattern (c): cooperators and
defectors are evenly mixed. Here we will focus on the
strategy transition between the circled red cooperator and the
circled blue defector. Before the formal analysis, it is indispensable
to predefine some useful parameters. We assume that
$f^a_C, f^b_C, f^c_C$ ($f^a_D, f^b_D, f^c_D$) are the fitness of the
circled cooperator (defector) in pattern (a), pattern (b), pattern
(c), respectively;  $\Delta_a, \Delta_b, \Delta_c$ are the
fitness difference of the circled cooperator and the circled
defector in pattern (a), pattern (b), pattern (c), respectively; and
$W^a_{C \rightarrow D}, W^b_{C \rightarrow D}, W^c_{C \rightarrow
D}$ ($W^a_{D \rightarrow C}, W^b_{D \rightarrow C}, W^c_{D
\rightarrow C}$ ) are the transition possibility of the circled
cooperator to become defector (the circled defector to become
cooperator) in pattern (a), pattern (b), pattern (c), respectively.
According to Eq.(2), we have:

\begin{equation}
\begin{array}{l}
\label{abcc} {\Delta_a} = f^a_D- f^a_C = (4b-3)+(3.75-5b){u}\\
 \\{\Delta_b} = f^b_D- f^b_C = b(1-2{u})\\
 \\{\Delta_c} = f^c_D- f^c_C = 2(b-1)(1-{u}).
\end{array}
\end{equation}

The fitness difference of cooperators and defectors is crucial for the
evolution of cooperation. Several previous works have shown that the
emergence of cooperation on lattices is often induced by the
formation of cooperator clusters, where cooperators can obtain
higher payoff (fitness) to protect themselves against the invasion
of defectors. From Eq.(4), we can find
${\Delta_a},{\Delta_b},{\Delta_c}$ monotonously decrease with the
increase of $u$, which indicates large value of $u$ can weaken the
advantage of defectors, and will be greatly beneficial for the
evolution of cooperation. Additionally, Figure 4 clearly demonstrates the
transition possibility of the circled cooperator and circled defector when varying
value of $b$ for three sub-patterns. One can observe that, although the
cross points of $W_{D \rightarrow C}$ and $W_{C \rightarrow D}$ are different, the result is
consistent that as $u$ increases,  a defector is more likely to turn into a cooperator yet
a cooperator is harder to become a defector.

\begin{figure}[htbp]
\begin{center}
\scalebox{0.3}[0.3]{\includegraphics{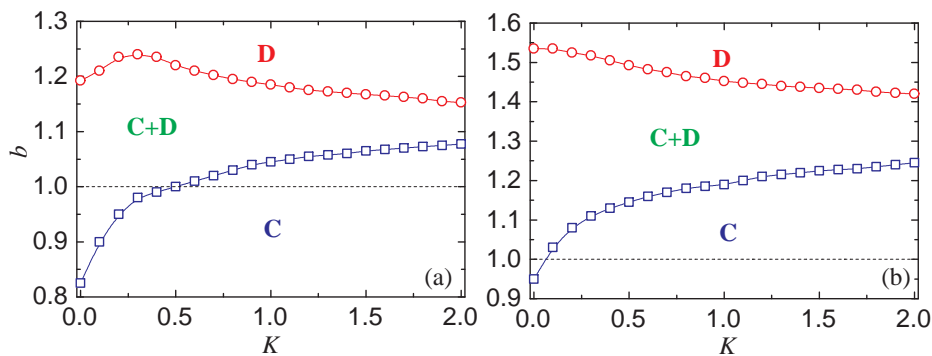}}
\caption{(\textit{color online})Full $b-K$ phase diagram for $u=0.2$
(a) and $u=0.4$ (b), obtained via Monte Carlo simulations of the
prisoner's dilemma game on the square lattice. The blue and red
lines mark the border between stationary pure C and D phases and the
mixed C+D phase, respectively. Resembling with previous works
\cite{20,22}, it can be observed that for $u=0.2$ (a) there exists
an intermediate uncertainty in the strategy adoption process (an
intermediate value of $K$) for which the survivability of
cooperators is maximal, \textit{i.e.}, $F_{\rm c}$ is maximal.
However, the borderline separating the pure C and the mixed C+D
phase for the $u=0.4$ case (b) exhibits a monotonically decreasing
curve, the D $\leftrightarrow$ C+D transition is qualitatively
different.}
\end{center}
\end{figure}


Furthermore, we will examine how the parameter $u$ affect the evolution
of cooperation under different levels of uncertainty. The later can be
tuned via $K$ (see Eq.(3)),which acts as a temperature parameter in the employed
Fermi strategy adoption function \cite{11}. In the case of $K \to 0$, the strategy
of selected neighbor is always adopted provided that its fitness is higher. While
in the limit $K \to \infty$, all information is lost, and switching to neighbor's
strategy is like tossing a coin. For accurately solving the problem, we obtain the full
$b-K$ phase diagrams for some characteristic values of $u$ on the square lattice.

The phase diagram illustrated in Fig.5(a) is well-known, and implies the existence of
an optimal level of uncertainty for the evolution cooperation, as was previously reported
in \cite{20,22}. In particular, note the the D $\leftrightarrow$ C+D transition line is bell shaped,
indicating that $K \approx 0.3$ is the optimal temperature where cooperators are able to survive
at the highest value of $b$. This phenomenon as the evolution resonance can only
be observed on the interaction topologies lacking the overlapping triangles \cite{19}.
On the other hand, comparing to the traditional phase diagram \cite{20,22,32}, {{\it i.e.} $u=0$},
the survival space of mixed strategies C+D is enhanced greatly, which will be helpful
for the emergence of cooperation. Interestingly, the increasing of $u$ completely eradicates
(as do interaction networks incorporating overlapping triangles) the existence of an optimal
$K$, as presented in Fig.5(b). The bell shaped D $\leftrightarrow$ C+D transition line gives way to a
monotonously decreasing line, indicating the enhancement of uncertainty level will directly accelerate
the dying out of cooperators. This quantitative change in the phase diagram implies that increasing
the tunable parameter $u$ or the proportion of environment in the individual fitness
will affectively alter the interaction networks. Though the square lattice evidently lacks the overlapping
triangles, which guarantees the existence of an optimal $K$ for the small values of $u$,
considering neighborhoods in the fitness seems to alter the likelihood and enables the linkage of essentially disconnected
triplets realized. A similar report was recently investigated in public goods games, in which the joint membership in large
groups was also found to alter the effective interaction network and thus the impact
of uncertainly on the evolution of cooperation \cite{46}.

\begin{figure}[htbp]
\begin{center}
\scalebox{0.25}[0.25]{\includegraphics{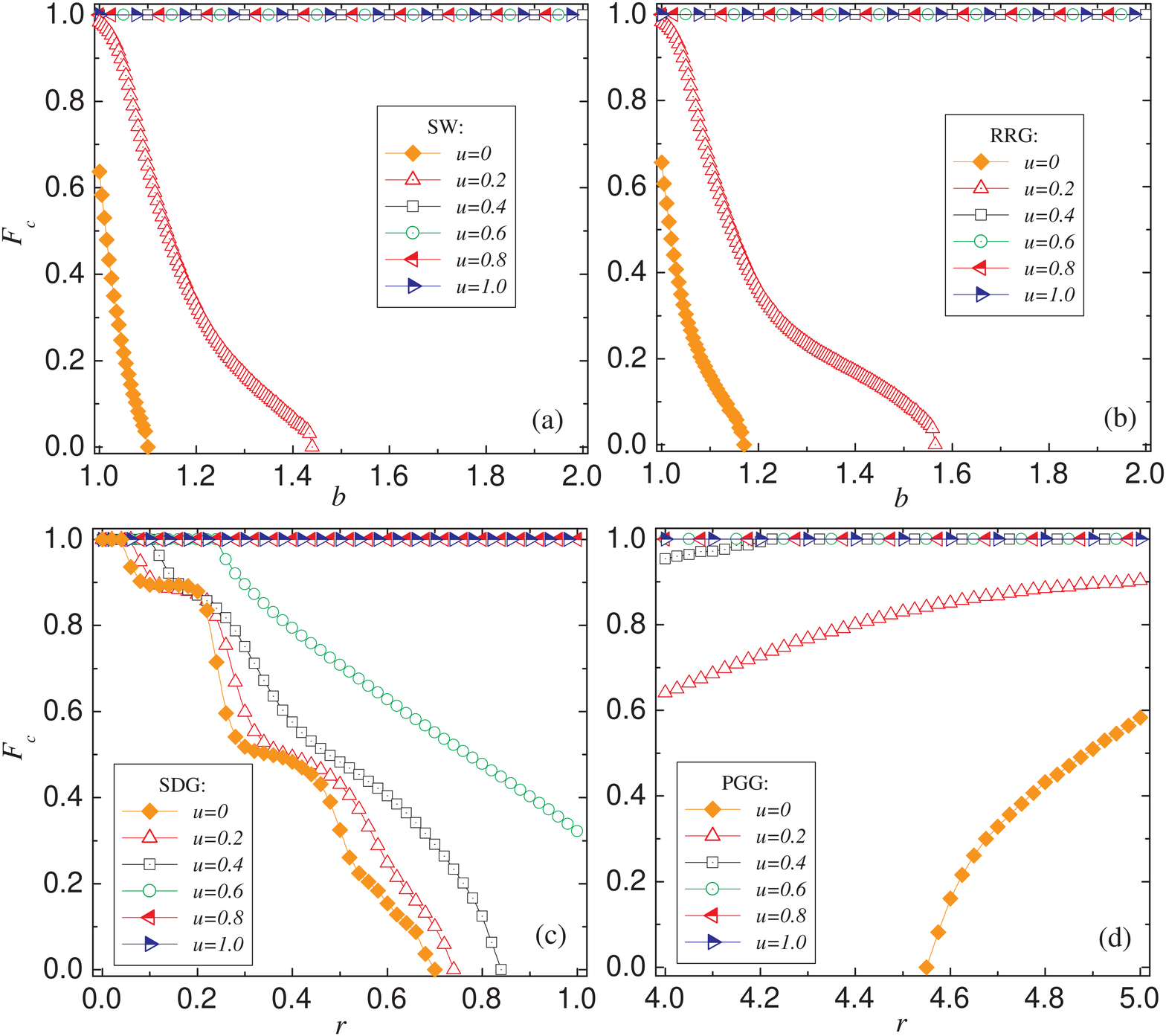}}
\caption{(\textit{color online}) Frequency of cooperators ${\it
F_{\rm c}}$ in dependence on the parameter $b$ (or $r$) for
different values of the parameter $u$ for different complex networks
and different game models: (a) prisoner's dilemma game on the
Watts-Strogatz small world network (SW) with the fraction of rewired
links equalling 0.1; (b) prisoner's dilemma game on the random
regular graph (RRG); (c) snowdrift game on the regular lattice; (d)
public goods games on the regular lattice. Note that these results
are in qualitative agreement with those obtained in Fig.2(a) in that
the larger value of $u$ greatly facilitates cooperation. Depicted
results were obtained for $K=0.1$.}
\end{center}
\end{figure}

Lastly, it remains of interest to explore the generalization
of cooperation promotion for such a new mechanism, we investigate
the systematic cooperative behaviors under different complex networks and
different game models (Fig.6). Similarly as Figure 2, it can be
observed that the cooperation level can be greatly promoted with the
increzse of $u$ for prisoner's dilemma game on the small world
network and the random regular graph. At the same time, it shows that
the evolution of cooperation can be effectively guaranteed for snowdrift game and public
goods game on square lattices. The result is specially inspiring in snowdrift
game, since cooperation is usually thought to be inhibited by the spatial
structure \cite{36}.

\section{Conclusion}
In summary, we have investigated the effect of redefined fitness
on the cooperative behaviors whin the framework of spatial
evolutionary prisoner's dilemma game. In the model, the fitness of
an individual is evaluated by the combination of its current payoff
and environment (the average value of all its neighbors' payoffs). It
has been found that integrating neighborhoods in the evaluation of fitness can promote a
remarkably high cooperation level, especially when it
holds larger proportion in the fitness. Via exploring some
typical sub-patterns in spatial games, we have demonstrated that cooperators
considering the influence of environment can better resist the invasion of
defectors.

Further interesting is the fact the increase of the parameter $u$,
marking more proportion of environment, has altered the effective interaction
network, while the similar phenomenon can be observed for public goods games in \cite{46}.
This effective transition of topology structure will provide more
beneficial condition to cooperators. To prove the generality of the promotion effect,
we have also examined the cooperative behaviors for different neighborhood sizes, different
topology structures and different game models. It is instructive that the similar results can be
observed. Our work reveals that considering environment {\it i.e.}, integrating
neighborhoods in the evaluation of fitness, plays an important role in the
evolution of cooperation, and thus it may shed a light on understanding
the emergence of cooperative behaviors in natural and social systems.

\section*{Acknowledgements}

Wang and Zhang thanks the support by the Center for Asia Studies of
Nankai University (Grant No. 2010-5), the National Natural Science
Foundation of China (Grant No. 10672081). Du and Cao thanks the
support by the National Basic Research Program of China (Grant
No.2011CB707000) and the Foundation for Innovative Research Groups
of the National Natural Science Foundation of China (Grant No.
60921001). This work has benefited substantially from the
insightful comments of the referees, and we
appreciate their help.





\end{document}